\documentclass[10pt,twocolumn]{article} 
\usepackage{simpleConference}
\usepackage{times}
\usepackage{graphicx}
\usepackage{amssymb}
\usepackage{url,hyperref}
\usepackage{subfigure}

\begin{document}

\title{Objective-oriented method for uniformation of various directivity representations}

\author{Adam Szwajcowski \\
\\
AGH University of Science and Technology \\ Department of Robotics and Mechatronics
\\
szwajcowski@agh.edu.pl  \\
}

\maketitle
\thispagestyle{empty}

\begin{abstract}
Over recent years, numerous attempts were taken to provide efficient methods of directivity representation, either regarding sound sources or head-related transfer functions. Because of the wide variety of programming tools and scripts used by different researchers, the resulting representations are inconvevnient to reproduce and compare with each other, hampering the development of the subject. Within this paper, an objective-oriented method is proposed to deal with this issue. The suggested approach bases on defining classes for different directivity models that share some general properties of directivity functions, allowing for easy comparison between different representations. A basic Matlab toolbox utlizing this method is presented alongside exemplary implementations of directivity models based on spherical and hyperspherical harmonics.
\end{abstract}

\section{Introduction}
\label{s:introduction}

Directivity functions are one of the most complex characteristics of sound sources and receivers, as they describe either ability to emit sound or sensitivity depending on multiple variables, most notably on the direction, but also on the frequency and on the distance from the measured object. The directivity data can be measured at a finite number of combinations of these variables, which makes the measurement results merely a set of discrete samples of a function that is fundamentally continuous over four different dimensions\footnote{Direction definition requires two dimensions (typically two angles).}. Because of the complexity and the unique nature of directivity, researchers all over the world used slightly different formats to store the measurement data, often in the form of Matlab files of different structures \cite{Gardner1995}\cite{Algazi2001}. In 2015, Majdak et al. came up with Spatially Oriented Format for Acoustics (SOFA), which standarized storing various spatial characteristics such as sound source directivity or head-related transfer functions (HRTFs) \cite{Majdak2013}\cite{SOFA}. Introduction of SOFA made accessing measurement data much more intuitive, yet still non-trivial, since the format leaves some freedom as far as units and coordinate systems are concerned; the files might be inconvenient to read and process automatically as the responsibility for unification of data definition is moved from the creator to the user\footnote{Although most files use the default values of units and coordinate systems, it is not always the case and thus has to be checked to be sure that the data is read properly}.

It is important to acknowledge, that directivity data can be stored not only as a set of discrete measurement values. In fact, over the years, multiple methods of directivity representation were suggested (especially regarding HRTFs), most notably spatially continuous models based on spherical harmonics (SHs), introduced by Evans et al. in 1998 \cite{Evans1998}. Even then, the SH representation is a whole family of directivity models incorporating different preprocessing techniques to maximize accuracy and efficiency \cite{Brinkmann2018}. On top of that, there are many other models utilizing e.g. filter definitions \cite{Kulkarni1995}\cite{Kulkarni2004}, spherical wavelets \cite{Hu2019} or different basis functions \cite{Zhang2009}, to name just a few. Not only the models differ significantly, but also different authors used different measures to verify the accuracy of their solutions. To reliably compare different methods or representations, one would have to reproduce the computations. Even having obtained all the source code from the researchers, this is no easy task, since the data can be loaded and saved after processing in a wide variety of ways. Hence arises a need for some sort of a uniformation of various directivity representations.

This paper explains a novel objective-oriented approach to storing directivity models and discusses key features of its basic implementation. In section \ref{s:assumptions}, the assumptions and goals for the method are laid out. Section \ref{s:implementation} presents a Matlab toolbox designed basing on the proposed method. Section \ref{s:examples} provides examples of directivity representations built on top of the toolbox. The current version and possible improvements are discussed in section \ref{s:discussion}. Section \ref{s:summary} briefly concludes the work.

\section{Assumptions}
\label{s:assumptions}

When designing a function library that could handle different directivity representations, the first and foremost assumption was that the framework has to be flexible enough to support various models. The main focus was put on adapting the library for approximations utilizing basis functions. Such representations consist of sets of coefficients and can be continuous over some dimensions and discrete over the others. It is thus clear, that different models will need different data structures to store their coefficients and special care has to be taken to support any possible directivity representation.

Despite the flexibility assumption, all directivity models do have certain common features. It is then important to provide a uniform interface, that will take advantage of the common properties, while leaving full freedom as far as model definition is concerned. For example, a directivity object defined in any representation must be able to provide its values for queried coordinates and possibly use them to plot spectra or directivity ballons. However, due to mentioned variability of models, sometimes it might be impossible to read the data precisely at requested positions and the library should be equipped to tackle such problems.

Another important feature to be included would be modularity. Even currently, there are many different directivity models and next ones are sure to come in the future. It is thus desirable to design the library in such a way that new representations can be easily added and shared between researchers. The modularity is linked with compactness, which is the last one of the key features requiered from the library; closing all the source code related to a given representation within a single file would make the overall user experience much better as far as managing directivity models and objects is concerned. 

\section{Implementation}
\label{s:implementation}

The library utilizing objective-oriented approach was written in the form of a Matlab toolbox under the name \textit{ooDir}, an abbreviation of Objective-Oriented Directivity. Current version of the toolbox can be found at \url{https://sourceforge.net/projects/oodir}. After carefuly reflecting upon requirements set for the task, a simple and compact structure of the library was developed with only three classes constituting the core of the toolbox:

\begin{enumerate}
\item{Directivity - superclass from which all the classes for various directivity representations inherit a common interface.}
\item{Coordinates - class containing and managing data coordinates.}
\item{RawDirectivity - Directivity subclass for raw measurement data, the basis for every other representation.}
\end{enumerate}

\begin{figure}
\centering
\includegraphics[width = \linewidth]{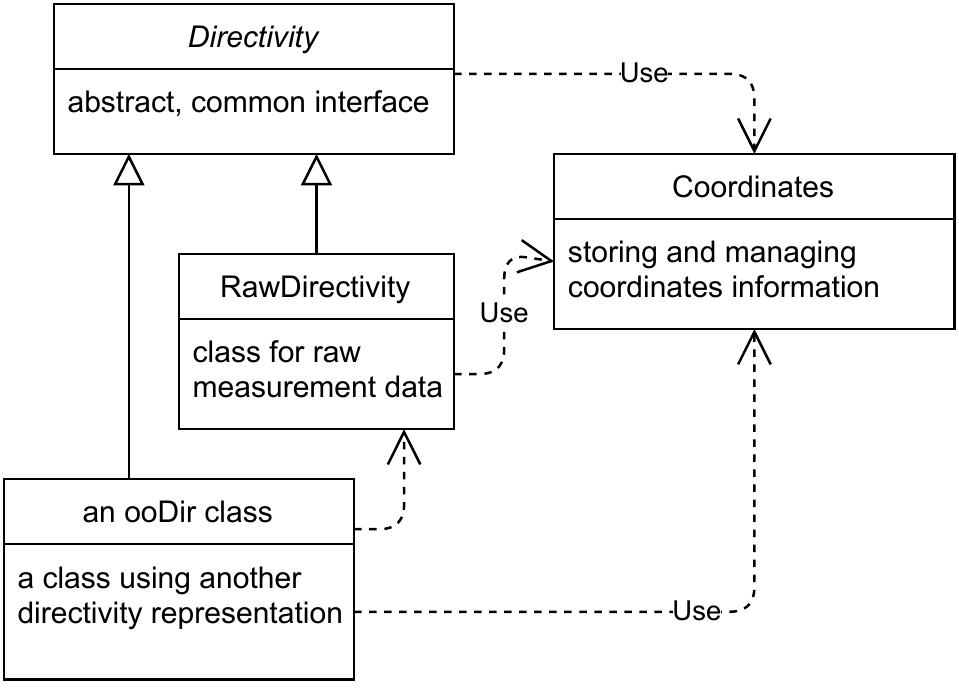}
\caption{Class diagram of the core classes and an exemplary class built on top of them. Objects for other directivity representation classes are built based on corresponding RawDirectivity objects.}
\label{f:structure}
\end{figure}

On top of these three basic classes, other classes for different directivity representation can be written. An object of a Directivity subclass contains full information on a directivity function expressed using certain representation method coded withing the subclass. The relations between the classes are shown in Fig.~\ref{f:structure}. In the following subsections, the structure and key features of each of the three core classes will be described in more details.

\subsection{Directivity class}
\label{s:directivity}

Directivity is an abstract class defining the structure of inheriting classes containing certain directivity models. To maintain flexibility of these models, the Directivity class contains only two properties that are common for every directivity representation; one of them is \texttt{info}, a string for general information about given obejct and the other is \texttt{continuity}, a three-element vector of booleans describing whether given representation is continuous along direction, frequency and distance, respectively. \texttt{continuity} is predefined for each of the subclasses.

Methods of the Directivity class are meant to provide a common interface for reading data from objects. Again, this can vary depending on a given model, which is why core methods \texttt{GetDataM} and \texttt{GetDataCoords} are abstract. \texttt{GetDataCoords} simply return discrete coordinates for a given directivity object, while \texttt{GetDataM} requires a Coordinate input to determine the coordinates at which data are requested. The output of \texttt{GetDataM} has a clear structure - it is a {3-dimensional matrix}, where consecutive dimensions are linked with directions, frequencies and distances, respectively. Basing on this structure, Directivity provides methods for plotting spectra and directivity balloons, both in two variations depending on whether given model is discrete or continuous along freqency or space, respectively.

\subsection{Coordinates class}
\label{s:coordinates}

Coordinates is a simple class for storing and managing data coordinates. It has three properties: \texttt{dirs}, \texttt{freqs} and \texttt{dists} to store directions, frequencies and distances, respectively. The constructor of the Coordinates class simply passes its arguments to above-mentioned private properties. \texttt{dirs} is a two-column matrix, where the left column is the azimuth and the right column is the elevation, expressed in degrees. \texttt{freqs} and \texttt{dists} are vectors containing values in hertz and meters, respectively. In case some of the coordinates are not needed (e.g. because of continuity along that dimension), they are made empty matrices. Furthermore, due to directivity being usually expressed only in far-field, the entire toolbox is designed in such a way that the distance can be ignored and it will be assumed that it is 1 m.

Coordinates also includes coerce methods useful for reading data at discrete dimensions. If data are unavailable at requested coordinates, these functions coerce them to the closest available ones and return both the values and the coerced coordinates (direction, frequency or distance).

\subsection{RawDirectivity class}
\label{s:rawdirectivity}

Structuraly, RawDirectivity is one of many classes that can be built based on the Directivity and Coordinates classes; however, different directivity representations are always derived from raw measurement data and so a RawDirectivity object will serve as a basis for getting the directivity object of a different class, i.e. constructors for other classes will typically require a RawDirectivity object as their input. RawDirectivity simply provides a proper formatting of measurement data that can be read e.g. from a SOFA file.

As a subclass of Directivity, RawDirectivity is required to have two properties; \texttt{info} that can be provided as an input in the constructor and \texttt{continuity}, which is set in this case to \texttt{[0, 0, 0]} since measurement data are discrete over all considered dimensions. Furthermore, it contains a Coordinates object \texttt{coords} storing information on the coordinates at which the data were measured and a matrix with values at these coordinates \texttt{dataM} of the same structure as in Directivity's method \texttt{GetDataM}. It was decided, that the data would be stored in the form of logarithmic magnitude values. Even though there exist a variety of models using complex-frequency or even time representations, it is currently a more popular approach to focus on the perceptual significance of data by ignoring the phase and using logarithmic scale for the magnitude \cite{Romigh2015}\cite{Hartung1999}. However, if needed, the toolbox can be quickly adjusted for complex representations simply by changing the type od data from double-precision to complex\footnote{Directivity's methods for plotting spectra and balloons would also have to be adjusted, should anybody wanted to use them with complex data.}.

RawDirectivity has only three simple methods. One of them is the constructor, which simply passes its arguments to the properties. These arguments/properties are a description string \texttt{info}, a Coordinates object \texttt{coords} and a data matrix \texttt{dataM}. The other two methods are \texttt{GetDataM} and \texttt{GetDataCoords} inherited from Directivity. Both are meant to return private properties of a RawDirectivity objects and \texttt{GetDataM} also performs coordinate coercion if data are not available at the requested coordinates.

\section{Examples}
\label{s:examples}

The main reason behind the creation of the library was to provide tools for creating other classes for different directivity representations, so that they would all have a common interface. In the following subsections, two examples developed by the author are presented to showcase various implementary aspects of such models/classes. Both these classes can be found at the ooDir repository (\url{https://sourceforge.net/projects/oodir}).

\subsection{Spherical harmonic approximation}
\label{s:SHA}

The first model, coded within the class ooDir\_SH, uses least-squares fitting of real SHs, without any regularization\footnote{Details can be found in \cite{Szwajcowski2021a}.}. The most prominent difference when compared to RawDirectivity is that the data matrix is replaced by a matrix of SH coefficients. Furthermore, in the properties there are also \texttt{lmax} and \texttt{mmax} representing maximal degree and order of SHs used for approximation. Finally, there is a \texttt{minElev} parameter, that states what is the minimal elevation for which the data can be retrieved accurately. It is useful for handling HRTF objects which lack data for very low elevations due to the measurement limitations; for example, one of the best-known HRTF datasets of KEMAR contains measurements only for elevations higher than -40 degrees \cite{Gardner1995}. Providing such a piece of information is useful to know the range at which data can be reproduced reliably, e.g. Directivity method for plotting directivity balloons will ignore the directions of elevation below the value of \texttt{minElev}.

The class constructor is slightly more advanced than in RawDirectivity, since it has to compute the SH coefficients. The required arguments are just a RawDirectivity object \texttt{RD} and maximal degree and order of SHs. Optionally, one can also provide description (if not, it propagates from \texttt{RD}) and flag to show waitbar for computations (false by default). The constructor computes SH values at \texttt{RD}'s coordinates through a separate static function and then determines the coefficients by means of Matlab's least-squares fitting function \texttt{lsqminnorm}. The same function generating SHs can be then used to retrieve discrete values from the coefficients using \texttt{GetDataM} method inherited from Directivity.

\begin{figure}[t!]
\centering
\subfigure[]{
	\includegraphics[width=0.95\linewidth]{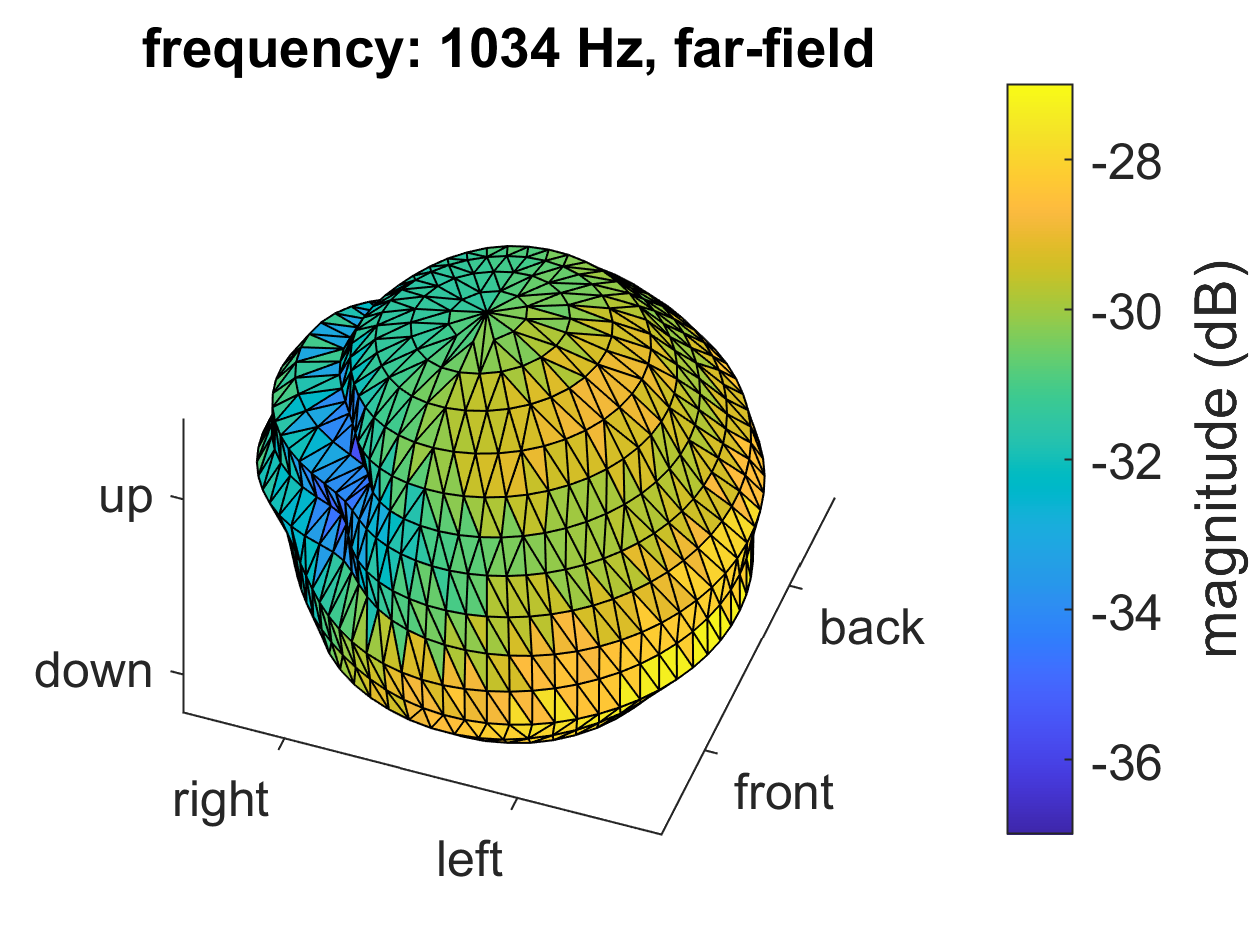}
	\label{f:RD}}
\\
\subfigure[]{
	\includegraphics[width=0.95\linewidth]{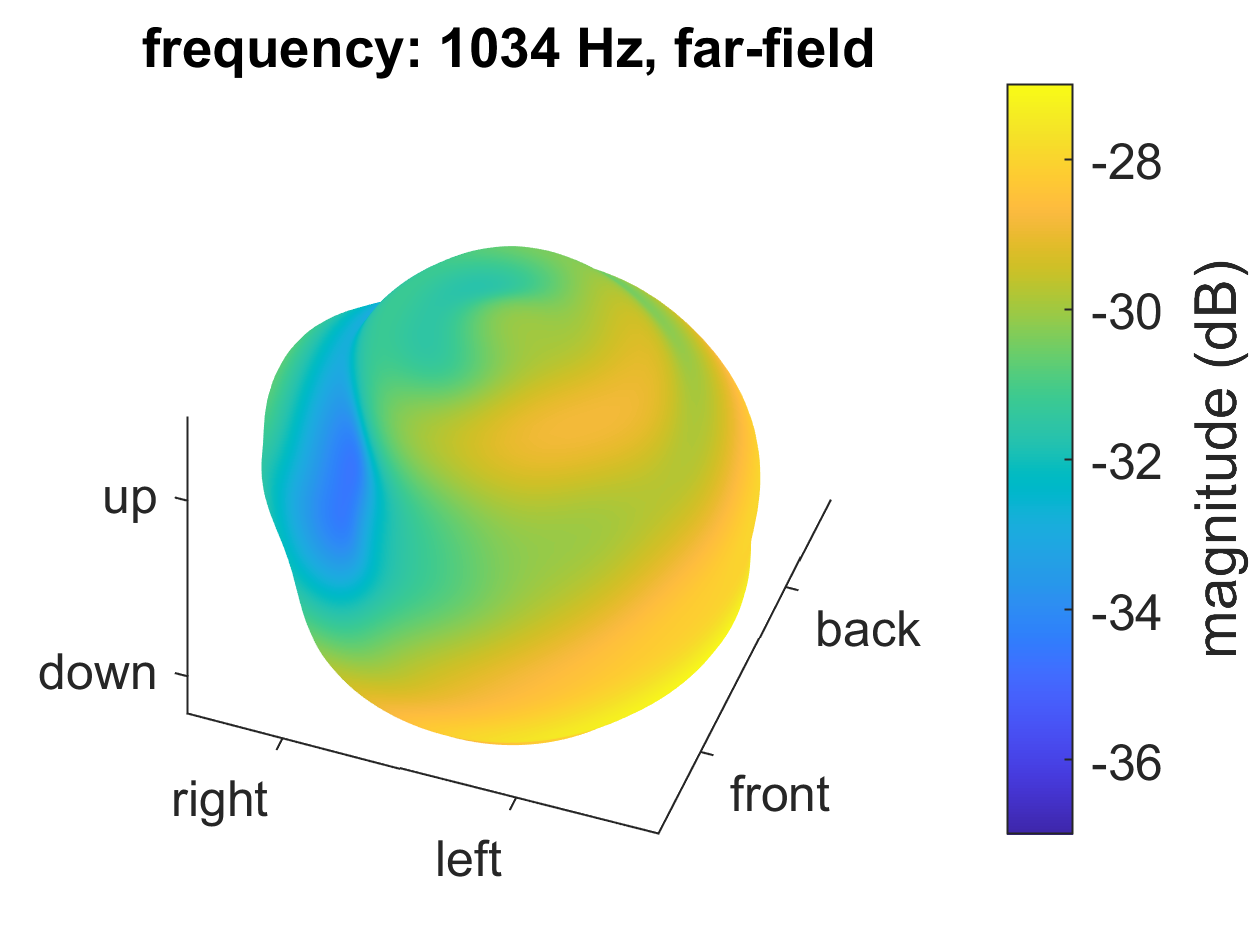}
	\label{f:SH}}
\caption{Directivity balloons plotted for RawDirectivity (up) and ooDir\_SH (down) objects built based on the original KEMAR HRTF measurement (left ear) \cite{Gardner1995}. The requested frequency was 1000 Hz, which was then coerced to the closest available frequency.}
\label{f:balloons}
\end{figure}

Data from RawDirectivity and ooDir\_SH objects can be accessed using the same functions; however, due to their different structures, different methods can be invoked in some cases, e.g. for a discrete object, all its datapoints will be plotted, while for continuous models, the datapoints will be generated with requested resolution (by default $2.5$ degrees for balloons and 10 Hz for spectra). Examples of two different variants for directivity balloons (discrete and continuous) are presented in Fig.~\ref{f:balloons}.
 
\subsection{Hyperspherical harmonic approximation}
\label{s:HSHA}
HSH approximation is a recently developed method of directivity representation providing continuity in both space and frequency domain. The \texttt{continuity} vector in class ooDir\_HSH is thus set to \texttt{[1,1,0]}, being only discrete in distance. Overall, the ooDir\_HSH class has a very similar structure to ooDir\_SH with just a couple of differences. As far as properties go, apart from adjusting the structure of the coefficient matrix and the number of basis function parameters, there is one addition - \texttt{maxFreq}, the maximum available frequency (usually equal to the Nyquist's frequency); since HSHs are continuous over frequency, the range has to be defined, in this case from 0 to \texttt{maxFreq}.

Furthermore, the HSH definition requires some external functions. Simplified formula for a HSH for given values of parameters $n$, $l$ and $m$ is following:

\begin{equation}
	Z_{nl}^{m}(\varphi,\theta,\psi) \equiv N(n,l)\sin^l{\psi} \; C_{n-l}^{l+1}(\cos{\psi}) \; Y_{l}^{m}(\varphi,\theta) \, ,
\label{e:HSH}
\end{equation}

where $N(n,l)$ is a normalization factor, $C_{\nu}^{\alpha}(x)$ are Gegenbauer polynomials and $Y_{l}^{m}(\varphi,\theta)$ are SHs. Since generating SHs was already implemented in ooDir\_SH as a static function, it can be easily reused within another class. For the same reason, Gegenbauer polynomials are defined in ooDir\_HSH as a static function as well\footnote{Matlab's implementation of Gegenbauer Polynomials is based on Symbolics Toolbox, making it very computationally inefficient. Instead, ooDir\_HSH utilizes a much faster, recursive definition.}, so that it can be reused in another ooDir class. The normalization factor on the other hand is unique to HSHs and hence was coded directly in the body of the HSH generating function.

\section{Discussion}
\label{s:discussion}

Presented implementation of proposed method of handling directivity models satisfies assumed goals. Described class structure supports wide variety of directivity representation, while maintaining a convenient, uniform interface. However, even though the presented basic version of the toolbox is already functionable and fits intended role, there are some dillemas associated with its design. One of the difficulties was finding the limit between flexibility and convenience; while the main focus was put on the former, there are some limitations that were put in place for the sake of smoother user experience. First limitation comes from using a 3-dimensional matrix to handle discrete directivity data (see section \ref{s:directivity} for more details). Such data structure could be a problem if a dataset of a nonuniform sampling was concerned, e.g. having different frequency resolution for different directions or different direction resolution for different distances. However, such datasets are extremely rare and thus the convenience was chosen over flexibility in this case\footnote{It is common in HRTF datasets that azimuth resolution is different for different elevations. Hence, azimuth and elevation are linked together and treated as a pair of angles to avoid problems with nonuniform spatial sampling.}.

Another limitation of the design is reliance on logarithmic magnitude values. The choice in this case was motivated by simplicity of the code. Constant changing between complex and logaritmic spectra would require some extra steps in many functions, which in the end might be not needed at all. However, if there was a need, a support for impulse responses or complex spectra could be implemented in future versions of the toolbox.

On the other hand, one of the greatest advantages of the designed class structure is its modularity. If researchers wanted to share source code for various directivity representations with each other, they can simply send a single Matlab file with a class containing implementation of their method\footnote{In some cases, when using external functions, there might be a need for sending more files or copying some static methods, e.g. ooDir\_HSH using SH generation from ooDir\_SH.} and optionally the RawDirectivity files that were used in the research. Such ease of file communication would be a significant step forward in the field as far as collaborations and reproducability of results are concerned.

\section{Summary}
\label{s:summary}

A novel method for handling directivity representations was proposed, with the main goal being improved experience of sharing and comparing various directivity models with each other. The method utilizes objective-oriented design in which each representation is a subclass of a common superclass interface. The paper describes both the concept and its basic implementation in details. The resulting Matlab toolbox is lightweight and easy to develop, as proven by two examples of functional models implemented on top of it. The proposed approach not only provides a convenient and elegant way to share source code of different representation between researchers, but also greatly empowers possiblity of reproduction and comparison of various models, pushing the entire field forward on the way to find optimal solutions for storing directivity characteristics.

\section*{Acknowledgements}

This research was supported by the National Science Centre, project No. 2020/37/N/ST2/00122.

\bibliographystyle{ieeetr}
\bibliography{biblio}
\end{document}